%% file: Paper.tex
\newcommand{\CLASSINPUTtoptextmargin}{2.4cm}
\newcommand{\CLASSINPUTbottomtextmargin}{4.6cm}
\documentclass[conference]{IEEEtran}
\IEEEoverridecommandlockouts
\usepackage{cite}
\usepackage{amsmath,amssymb,amsfonts}
\usepackage{algorithmic}
\usepackage{graphicx}
\usepackage{textcomp}
\usepackage{xcolor}
\usepackage{subfiles}
\usepackage{amsmath}

\def\BibTeX{{\rm B\kern-.05em{\sc i\kern-.025em b}\kern-.08em T\kern-.1667em\lower.7ex\hbox{E}\kern-.125emX}}

\begin{document}

% \title{Conference Paper Title*\\
% {\footnotesize \textsuperscript{*}Note: Sub-titles are not captured in Xplore and should not be used}

\title{Software Product Line for Metaverse: Preliminary Results}

\author{
    \IEEEauthorblockN{Filipe Fernandes}
    \IEEEauthorblockA{\textit{System Engineering and Computer Science Department} \\
    \textit{Universidade Federal do Rio de Janeiro – COPPE/UFRJ}\\
    Rio de Janeiro, Brazil \\
    ffernandes@cos.ufrj.br}
    
    \and

    \IEEEauthorblockN{Cláudia Werner}
    \IEEEauthorblockA{\textit{System Engineering and Computer Science Department} \\
    \textit{Universidade Federal do Rio de Janeiro – COPPE/UFRJ}\\
    Rio de Janeiro, Brazil \\
    werner@cos.ufrj.br}
}

\maketitle

\begin{abstract}
The Metaverse is a network of eXtended Reality applications (XR apps) connected to each other, over the Internet infrastructure, allowing network users, systems, and devices to access them. It is very challenging to implement solutions for XR apps, due to the combination of complex concerns that should be addressed: multiple users with non-traditional input and output devices, different hardware platforms that should be addressed, forceful interactive rates, and experimental interaction techniques, among other issues. Therefore, this work aims to present a Software Product Line (SPL)-based approach to support the development of Web XR apps. More specifically, we define a features model that represents similarities and variables (domain analysis); we defined a core composed of generic and reusable software artifacts (domain project); and we developed an interface to support the instantiation of a Web XR app family, named MetaSee Features Model Editor (domain implementation). This approach integrates with a component of the MetaSEE architecture (Metaverse for Software Engineering Education). A preliminary assessment found that Features Model has conceptual consistency from the point of view of the complexity of Web XR Apps multimodal interaction. As future work, features model and artifacts will be increased with improvements and an evaluation with a significant number of participants will be made.
\end{abstract}


% focus on multimodal interactions complexity of Metaverse
\subfile{Sections/Introduction}

% to talk about domain and application engineering 
\subfile{Sections/Background}
\subfile{Sections/RelatedWork}

\subfile{Sections/MetaSEEoverview}

% to present overview of my architecture AND the features model
\subfile{Sections/Approach}
\subfile{Sections/PreliminaryEvaluation}

% evaluation discussion
% \subfile{Sections/Discussion}

\subfile{Sections/Conclusion}

\section*{Acknowledgment}
The authors would like to thank CNPq and CAPES Brazilian Agencies for financial support.

\bibliographystyle{IEEEtran}
\bibliography{references-zotero}

% \appendix
% \chapter{Teste}

% \begin{figure*}[!h]
%     \centering
%     \includegraphics[scale=1]{Figures/screens.png}
%     \caption{Relationship between features and generated source code}
%     \label{fig:screens}
% \end{figure*}

\end{document}

%% file: Sections/Introduction.tex
\section{Introduction} \label{sec:intro}
After Facebook announced the corporate name change in October 2021, unveiling plans to spend tens of billions of dollars on what Zuckerberg called ``the next chapter of social connection", the Metaverse \cite{wang_survey_2022}. Metaverse can also be defined as ``a massively scaled and interoperable network of real-time rendered 3D virtual worlds that can be experienced synchronously and persistently by an effectively unlimited number of users with an individual sense of presence, and with continuity of data, such as identity, history, entitlements, objects, communications, and payments" \cite{ball_metaverse_2022}.

In general, the Metaverse is a network of eXtended Reality applications (XR apps) connected to each other, over the Internet infrastructure, allowing network users, systems, and devices to access them \cite{fernandes_towards_2022}. XR apps offer new solutions to futuristic and existing challenges in several fields, such as games, education, oil exploration, automotive design, among others. They use non-traditional input and output devices to convey richer and more productive multimodal experiences than counterparts with standard interfaces. Current XR experiences allow users not only to see but also touch, hear, and smell very complex 3D scenarios at interactive frame rates \cite{figueroa_intml_2009}. It is very challenging to implement solutions for these type of applications, due to the combination of complex concerns that should be addressed: multiple users with non-traditional input and output devices, different hardware platforms that should be addressed, forceful interactive rates, experimental interaction techniques, among other issues \cite{figueroa_intml_2009}.

Therefore, we are interested in applying Software Product Line (SPL) to the Metaverse. More specifically, we believe XR is a challenging field for these technologies due to the huge amount of variability, compared to the one present in traditional devices. Additionally, the use of SPL could create a more uniform and easy environment for XR developers.

As a result, we have defined an SPL-based approach to developing Web XR apps. It consists of three phases: domain analysis, domain design, and domain implementation. The first phase defined a features model, it represents the features of an XR app family, their commonalities and variabilities, and the relationships among them. In the second phase, a core of generic and reusable software artifacts based on the A-Frame as Domain Specific Language (DSL) \cite{gill_aframe_2017} was conceived. Finally, an interface prototype for instantiating an XR app family was developed. The proposed approach is developed in a Metaverse for Software Engineering Education research project, named MetaSEE \cite{fernandes_systematic_2022}. MetaSEE is an architectural proposal that aims to integrate SEE virtual worlds in order to provide better engagement in the SE teaching and learning process through immersive experiences. Therefore, the instantiation interface described in this work, called MetaSEE Features Model Editor, will integrate with the \textit{Development Tools} component of the MetaSEE architecture.

In order to validate the features model conceived in the domain analysis phase, a preliminary evaluation was carried out with 5 participants. We were able to verify that the modeled features are able to represent the complexity of the interactions of the multimodal interfaces of Web XR apps. In addition, we identified opportunities for improvement both in the interface, as well as in the features model.

The present work is organized as follows: in Section \ref{sec:background} SPL definition and concepts are presented. The related works are discussed in Section \ref{sec:relatedwork}. A brief contextualization of MetaSEE research is presented in Section \ref{sec:metasee}. In Section \ref{sec:features-model} our SPL proposal for Web XR apps is presented in detail. The design and results of the preliminary evaluation are discussed in Section \ref{sec:evaluation}. Finally, Section \ref{sec:conclusion} provides some conclusions and future perspectives.

%% file: Sections/Background.tex
\section{Theoretical Background} \label{sec:background}
SPL's primary function is to generate specific products based on the reuse of an infrastructure that includes: software architecture, components, design patterns and planning methods.

According to Software Engineering Institute (SEI), a SPL is a set of software-intensive systems sharing a common, managed set of features that satisfy specific needs of a particular market or mission, and that are developed from a common set of core assets in a prescribed way \cite{northrop_seis_2002}. Core assets are the essence of a product line and represent configurable elements used to build derived applications.

The development of a SPL is divided into: Domain Engineering (DE) - process of identifying and organizing knowledge about a class of problems (domain) to support its description and solution in a systematic way; and Application Engineering (AE) phase in which applications are built within the specific domain, based on the reuse of artifacts from the base core, previously developed in ED \cite{clements_software_2002}.

During the product line development process, particular aspects of products can be highlighted. The variability concept refers to points in the core assets where it is necessary to differentiate individual characteristics of products, being represented with a feature model \cite{fernandes_feature_2008}. In order to model variability, it is necessary to represent all domain concepts and their relationships explicitly.

A feature model represents a domain and aims to make homogeneous the concepts among the participants involved in the process, such as users, domain specialists, and developers. It represents the features of a system family, their commonalities and variabilities, and the relationships among them. In addition, it has a high level of abstraction and is used as a starting point for the feature selection to new products instantiation.

An important relationship in a feature model is the dependence among features that defines when some features should be included in product due to the presence of other features. On the other hand, it may also define when some features should be removed from the product due to the presence of other features.

This kind of model has some important concepts. From the point of view of variability, this type of model can represent:

\begin{itemize}
    \item \textbf{Variation Point}: reflects the domain parameterization in an abstract manner and must be configurable through its variants;
    \item \textbf{Variants}: available functions/concepts that must necessarily be linked to a variation point, acting as configuration alternatives for their respective variation point; and
    \item \textbf{Invariable}: non-configurable element in the domain.
\end{itemize}

Additionally, from the point of view of optionality, a feature can be classified as:
\begin{itemize}
    \item \textbf{Optional}: element that may or may not belong to an application derived from a domain; and
    \item \textbf{Mandatory}: element that must be instantiated by all applications derived from a domain.
\end{itemize}

\begin{figure}[ht]
    \centering
    \includegraphics[scale=.3]{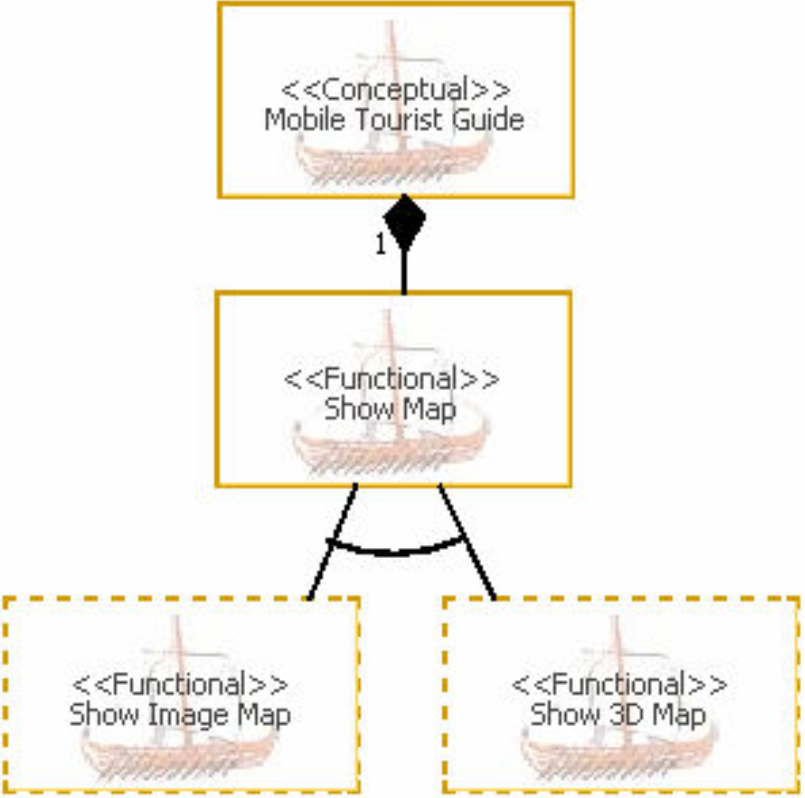}
    \caption{An example of features model \cite{zaina_uma_2015}}
    \label{fig:features-example}
\end{figure}

For example, considering the mobile tourist guide domain, we can have a functional feature ``show map" \cite{fernandes_feature_2008}. However, maps can be represented in more than one form, such as an image or a 3D map. In this case, we have a variation point with two variants (see Fig. \ref{fig:features-example}). However, we can also have a feature ``list hotels" that has the same behavior for all products derived from the product line. In this case, the ``list hotels" feature is considered an invariant.

%% file: Sections/RelatedWork.tex
\section{Related Work} \label{sec:relatedwork}
There have been previous work on SPL for XR apps. For example,  \cite{gottschalk_proconar_2020}
developed an approach for model-based AR-assisted product configuration based on the concept of Dynamic Software Product Lines. The authors present the corresponding tool support ProConAR in the form of a Product Modeler and a Product configurator. While the Product Modeler is an Angular web app that splits products up into atomic parts and saves it within a configuration model, the Product configurator is an Android client that uses the configuration model to place different product configurations within the environment of the customer. The possible products, which can be configured out of these parts, are stored in a feature model. At run time, this feature model can be  used  to  configure  3D  object  compositions  out  of  the  product  parts and adapt to user needs and environmental constraints. 

In this work, \cite{lebiedz_cave_2021} present the feature driven SPL solution based on that model allows for instantiation of different CAVE products based on the set of core assets and driven by a set of common Virtual Reality (VR) features with the minimal budget and time to market. A feature model specifies the requirements baseline for a given family of products, a set of core assets captures the underlying feature model and enable building the product from that set in a prescribed way, and production stations enable gradual instantiation of the application toward the final CAVE product.

In relation to the work of \cite{figueroa_intml_2009}, the authors applied software engineering concepts such as SPL, DSLs, and Model Driven Development (MDD) to the field of VR applications. The authors designed InTml as DSL with a very compact set of concepts that are common to any VR application: devices, behavior, and objects. InTml (Interaction Techniques Markup Language)\cite{figueroa_intml_2008} is an XML language for describing complex and implementation-independent VR applications. 

Despite the contributions of each work, we identified a gap in an approach that considers the range of interactions of XR apps, as well as being platform independent (e.g., desktop, mobile, CAVE). The work of \cite{gottschalk_proconar_2020} applies SPL to AR apps. The research developed by \cite{lebiedz_cave_2021} produces a family of applications based on Cave Automated Virtual Environment (CAVE). Finally, \cite{figueroa_intml_2009} uses InTml as a DSL to model the interaction features of VR apps.

Therefore, the contribution of this work aims to fill this identified gap. Our proposal is an SPL-based approach to the instantiation of families of Web XR apps. More specifically, we defined a set of reusable software artifacts in order to generate a skeleton code for Web XR apps, considering:

\begin{itemize}
    \item the diversity of devices, as well as their interaction and feedback specificities;
    \item applications with immersive experiences in both VR and Mixed Reality (MR); and
    \item application variability based on the two previous requirements.
\end{itemize}

%% file: Sections/MetaSEEoverview.tex
\section{Metaverse for Software Engineering Education} \label{sec:metasee}
MetaSEE is an approach to integrate SEE virtual worlds with the aim of filling a SEE gap identified through a systematic literature review \cite{fernandes_systematic_2022}. From the findings of the review, MetaSEE architecture was based on three fundamental requirements: \textit{SEE}, \textit{feasibility factors} and \textit{components and technologies}.

\textit{SEE} requirements establish the characteristics that the architecture must meet for the domain. They were based on the limitations of virtual worlds found by the review. Therefore, the architecture must: cover any SE topic; support the analysis of learning performance data; implement new forms of visualization and interaction; facilitate development for the Metaverse; perform analyzes using biometric data; and ensure data interoperability.

The Metaverse needs four factors to make it viable \cite{dionisio_3d_2013}. Therefore, the architecture must consider the following \textit{feasibility factors}:
\textit{realism} enables users to feel fully immersed in an alternative universe;
\textit{ubiquity} establishes access to the system via all existing digital devices and maintains the user's virtual identity throughout transitions within the system;
\textit{interoperability} is a connected collection of information, format, and data standards, most of which focus on the transfer of 3D models across virtual worlds, in addition to involving communication protocol, identity, and currency standards; and \textit{scalability} allows concurrent efficient use of the system by massive numbers of users.

\textit{Components and technologies} requirements establish the main aspects of the Metaverse identified through a comparative analysis of recent research \cite{fernandes_systematic_2022}: authoring tools; devices; economy; infrastructure; interaction; physical world; security; storage; technology; and virtual world.

\begin{figure}[ht]
    \centering
    \includegraphics[scale=.5]{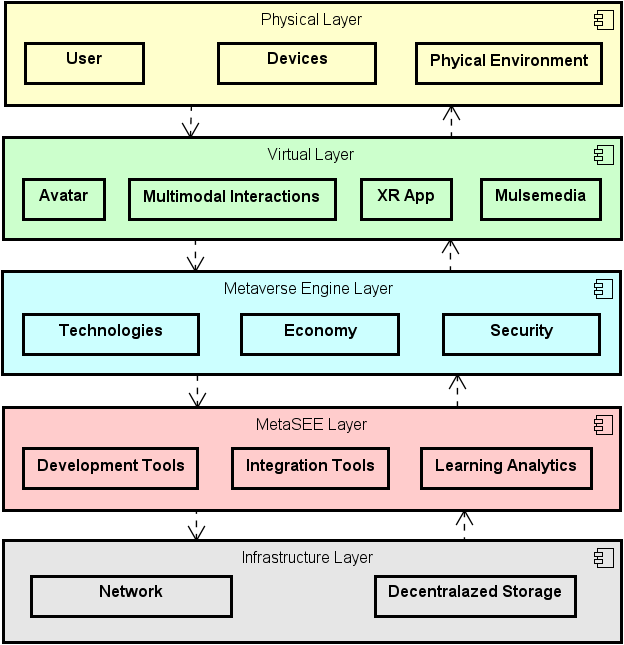}
    \caption{MetaSEE architecture overview \cite{fernandes_systematic_2022}}
    \label{fig:metasee-overview}
\end{figure}

As a result, MetaSEE architecture establishes the main elements to enable the Metaverse specifically between SEE virtual worlds. Fig. \ref{fig:metasee-overview} present the design of the MetaSEE and its main elements grouped in 5 layers.

\textit{Physical Layer} corresponds to the main entities external to the Metaverse and that belong to the physical and real world. \textit{Virtual Layer} establishes the main components of the ``virtualization" of physical layer elements. In \textit{Metaverse Engine Layer} establishes the general technologies, as well as economics and security of the Metaverse. \textit{MetaSEE Layer} is the main contribution to support SEE through the Metaverse. According to the established fundamental requirements, it was defined development tools, integration tools, and Learning Analytics (LA) as the components of this layer. Finally, \textit{Infrastructure Layer} deals with network and decentralization aspects of the Metaverse. 

In fact, MetaSEE is an overview of essential elements in order to specifically enable a Metaverse for SEE. Although its structure seems to serve any area with educational purposes, its main objective is to adopt a strategy of integrating XR apps for SEE and facilitating its adoption by the SEE community.

\begin{figure}[ht]
    \centering
    \includegraphics[scale=.3]{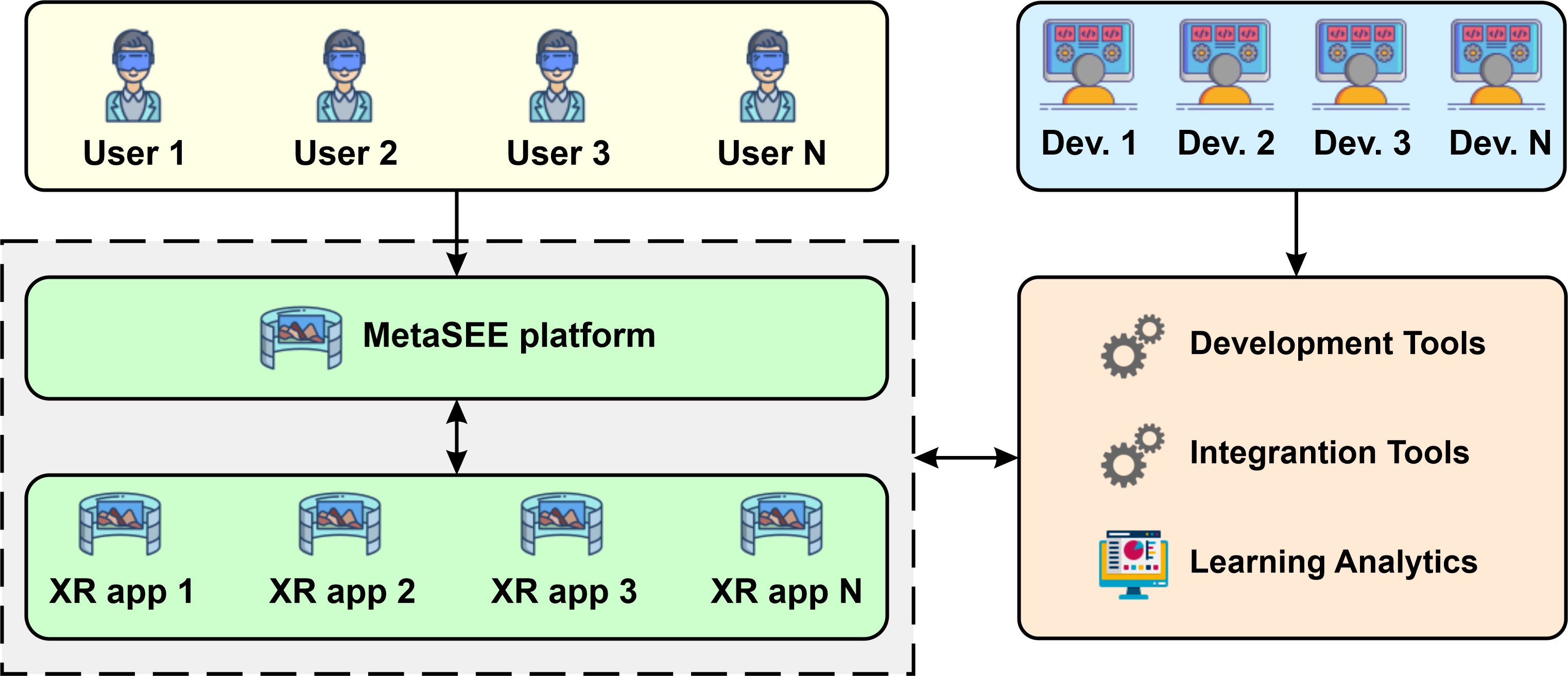}
    \caption{MetaSEE platform overview}
    \label{fig:metasee-platform}
\end{figure}

Therefore, as shown in Fig. \ref{fig:metasee-platform}, MetaSEE platform is an XR app that should make a link between users (e.g., students and teachers) and other XR apps for SEE. From the platform, the user will be able to choose which virtual world to ``enter" or ``exit", according to the approach and topic of SE addressed, in addition to ensuring data traceability for future analysis and improvement of teaching experiences. Additionally, developers will have access to development and integration tools serving as support in the reuse of features by XR apps, as well as in the development of new applications.

%% file: Sections/Approach.tex
\section{SPL-based Approach for Metaverse} \label{sec:features-model}
SPL-based approach for Metaverse is composed of the following phases: (i) domain analysis; (ii) domain project; and (iii) implementation of the domain. Domain analysis defines the scope of the SPL. The domain design describes the main elements in a generic way. The domain implementation provides model-based parameters to generate skeleton code to be reused by Web XR apps.

\subsection{Domain Analysis}
Domain analysis was performed on existing Web XR apps for defining the scope of the developed SPL. This step basically resulted in three products: (i) a list of general requirements for Web XR apps; (ii) the list of functionalities for the end user, classified according to presence in existing applications and possibility of reuse; and (iii) modeling the characteristics of the SPL, based on these requirements and functionalities, represented by a features model.

General requirements for existing Web XR apps are: (i) work in web browsers;
(ii) allow the use of traditional (e.g., desktop and smartphone) and non-traditional devices (e.g., XR headset and data glove); (iii) use the human senses to engage in immersive experiences; and (iv) allow various forms of interaction.

\begin{figure*}[ht]
    \centering
    \includegraphics[scale=.45]{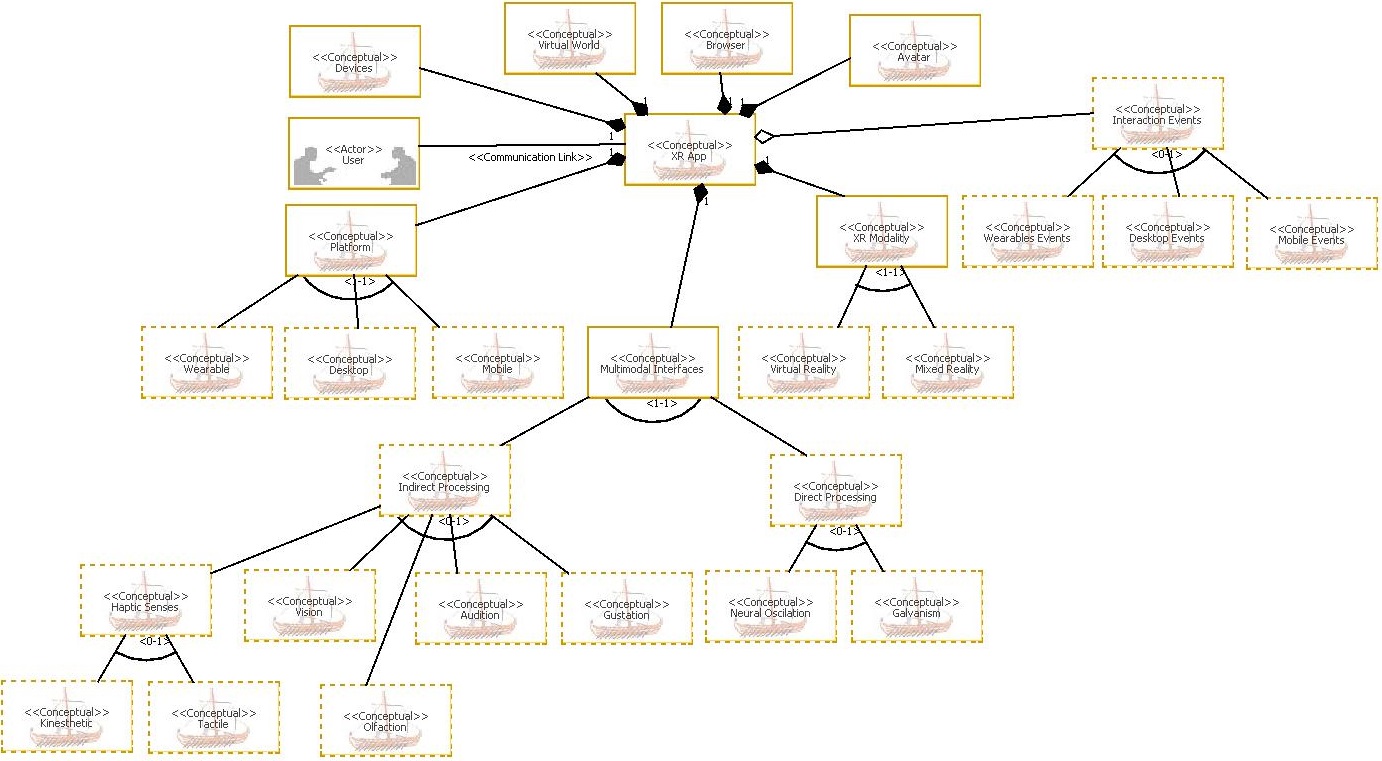}
    \caption{Features model for Web XR apps}
    \label{fig:features-model}
\end{figure*}

These requirements, together with the point functionalities, were classified in terms of presence in the applications. A common functionality is one that is or could be present in more than one family member, while a specific functionality is present in only one application and cannot be included in the others. This mapping resulted in the definition of the characteristics of the SPL, presented in Fig. \ref{fig:features-model}. For the modeling, the notation Odyssey-FEX \cite{blois_variability_2006} provided by the Odyssey environment \cite{braga_odyssey_1999} was used. Odyssey-FEX is an extension of the FODA method \cite{kang_feature-oriented_2002}, and provides a richer taxonomy for representing domain concepts and the relationships between them.

In the model, a Web XR app is described by the mandatory features \textit{Platform}, \textit{Multimodal Interfaces}, \textit{XR Modality}, \textit{Devices}, \textit{Browser}, \textit{Avatar}, and \textit{Virtual World}.

\textit{Platform} is a variation point that establishes variants of the type of device the application will run on. Wearable represents immersive devices (non-traditional), such as XR headset, haptic devices, motion sensors, etc. Desktop and Mobile represent the common interaction devices between the user and computer.

\textit{Multimodal Interfaces} is a variation point that establishes which modalities (i.e., human senses) that may be present in the applications in order to enable the engagement of immersive experiences. The modeling of this variability is based on the taxonomy of interaction modalities for XR \cite{augstein_human-centered_2019}. Direct processing works directly between a computer and the brain or muscles. Indirect processing refers to the multi-stage process where an output stimulus is perceived by a human receptor and then the information is delivered via electrical signals for further processing to the brain. The flow is similar for input stimulus from a human via sensors to the computer.

We based on Milgram’s Reality-Virtuality (RV) continuum \cite{milgram_augmented_1995} to define the \textit{XR Modality} variants. Virtual Reality considers only virtual objects to compose the application's virtual world and Mixed Reality combines real-world elements and virtual objects to compose the virtual world.

\textit{Devices} corresponds to the devices that will be used for both interaction and feedback (e.g. Meta Quest, HTC Vive, HoloLens); \textit{Browser} is the web browser that will be compatible with the application, \textit{Avatar} is the representation of the user, and \textit{Virtual World} is the virtual space in which the user is inserted to interact with virtual objects.

The only optional characteristic of modeling is \textit{Interaction Events}. It establishes interaction events for each type of device. For example, the implementation of the ``click" event considering the execution of an application on the desktop is different on a XR headset. In the desktop this event is fired through the mouse button, but in XR Headset can be through any button of yours controls. This characteristic is optional because an application can only display virtual objects without requiring any kind of interaction and/or feedback.

%No diagrama, uma iMA é descrito pelas características obrigatórias: todas as de Comunicação, de Configuração e de Gerenciamento de Atividades, apenas uma das Operações de Domínio, que representam as funcionalidades específicas do domínio de cada iMA (geometria, combinatória, etc.), e quais opcionais possui das categorias Funcionalidades sobre Atividades e Funcionalidades sobre Operações de Domínio.

\subsection{Domain Project}
The purpose of this phase is to specify a structure to be followed by applications from the modeled domain \cite{northrop_seis_2002}, i.e., software artifacts that belong to a particular domain and composed of a standard structure for the construction of applications.

For the project, we considered the feature model built in the previous phase and the documentation of the A-Frame framework \cite{a-frame_-frame_2022}. Therefore, the code skeleton generation will be based on this framework. The core of A-Frame is defined by the Entity-Component-System (ECS) architecture:
\begin{itemize}
    \item \textbf{Entities} are container objects into which components can be attached, and are represented by the \textit{a-entity} element and prototype;
    \item \textbf{Components} are reusable modules or data containers that can be attached to entities to provide appearance, behavior, and/or functionality, and  are represented by HTML attributes on \textit{a-entity}'s; and 
    \item \textbf{Systems} provide global scope, management, and services for classes of components, and are represented by \textit{a-scene}'s HTML attributes.
\end{itemize}

\begin{figure}[ht]
    \centering
    \includegraphics[scale=.13]{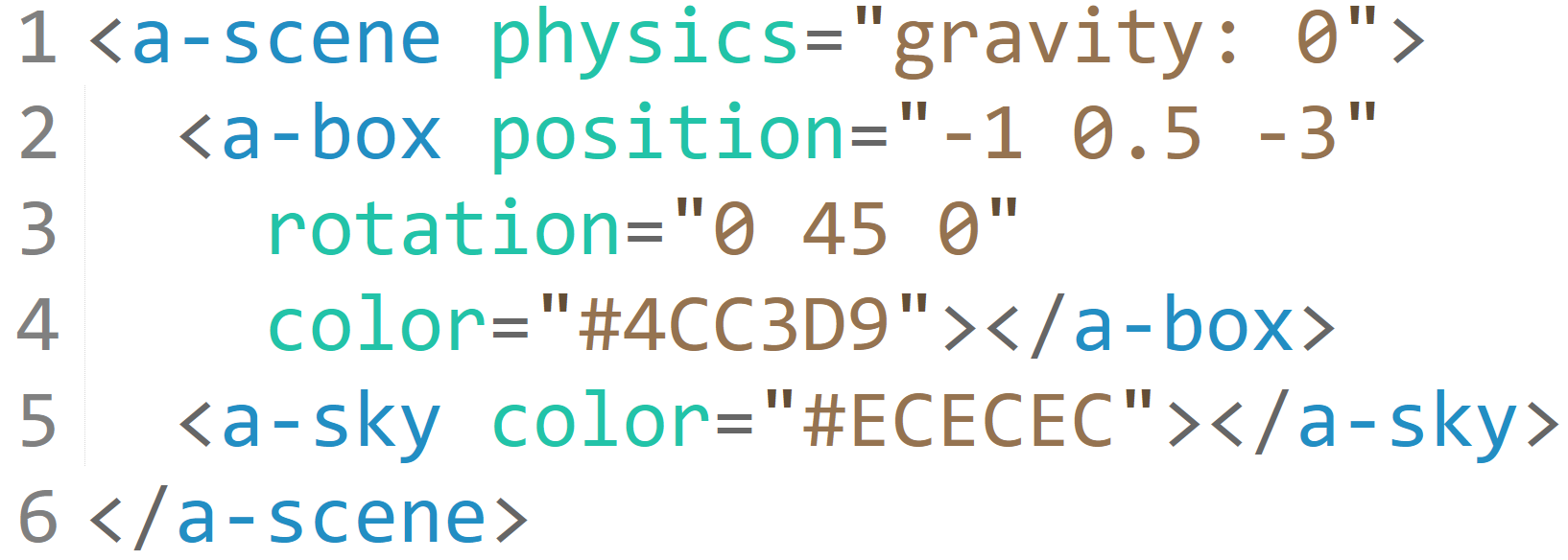}
    \caption{Example of A-Frame code}
    \label{fig:code-example}
\end{figure}

Fig. \ref{fig:code-example} shows an example of A-FRAME code. All the elements that will compose the virtual world must be contained in the entity \textit{a-scene}. Therefore, the entities \textit{a-box} and \textit{a-sky} are inside the tag \textit{a-scene}. These three tags are entities that are composed of components or systems. In line 1, \textit{a-scene} has the system \textit{physics} with its defined value as \textit{gravity:0}. From line 2, \textit{a-box} is composed of the components \textit{position}, \textit{rotation}, and \textit{color}. Both \textit{a-box} and \textit{a-sky}, defined in line 5, have the \textit{color} component, but with different values.

From its ECS Architecture, components can be developed by the community and implement specific features in order to reduce complexity in application development as well as promote code reuse. For example, \textit{aframe-event-set-component} provides a high level API to facilitate definition of mouse events or by gaze point. To change the behavior of an entity when clicking, the component \textit{event-set\_\_click} must be assigned.

As a result of this phase, we defined a comprehensive code template to define the inner workings of a Web XR app independently of the specific features of each application. Furthermore, we model the ECS architecture of the A-Frame based on the \textit{Composite} design pattern \cite{gamma_design_1995} in order to support the domain implementation.

\begin{figure}
    \centering
    \includegraphics[scale=.17]{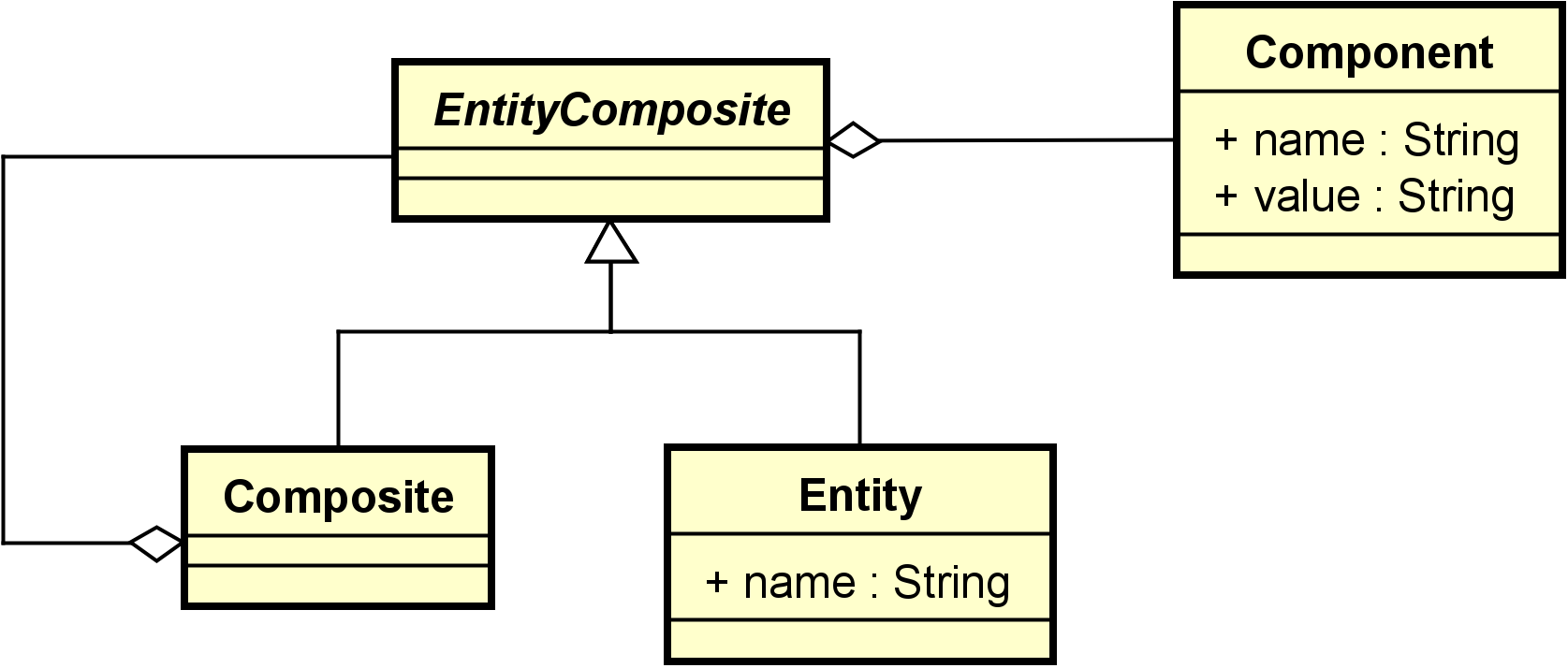}
    \caption{Abstraction of ECS architecture}
    \label{fig:composite}
\end{figure}

According to Fig. \ref{fig:composite}, an \textit{Entity} can contain others entities (\textit{Composite} class), and each \textit{Entity} can be composed by \textit{Components}. 

\subsection{Domain Implementation}
Finally, in this phase of domain implementation we build, through source code, reusable artifacts, based on modeling and domain design, as well as the interface for applications instantiation, named MetaSEE Features Model Editor. This tool consists of three components:

\begin{itemize}
    \item \textbf{Features}: comprises the modeled features in the domain analysis phase;
    \item \textbf{XR Component}: corresponds to A-Frame components that will be required for the implementation of features; and 
    \item \textbf{Template Generator}: generates the source code based on the selected features for the Web XR app instantiation.
\end{itemize}

\begin{figure}[ht]
    \centering
    \includegraphics[scale=.17]{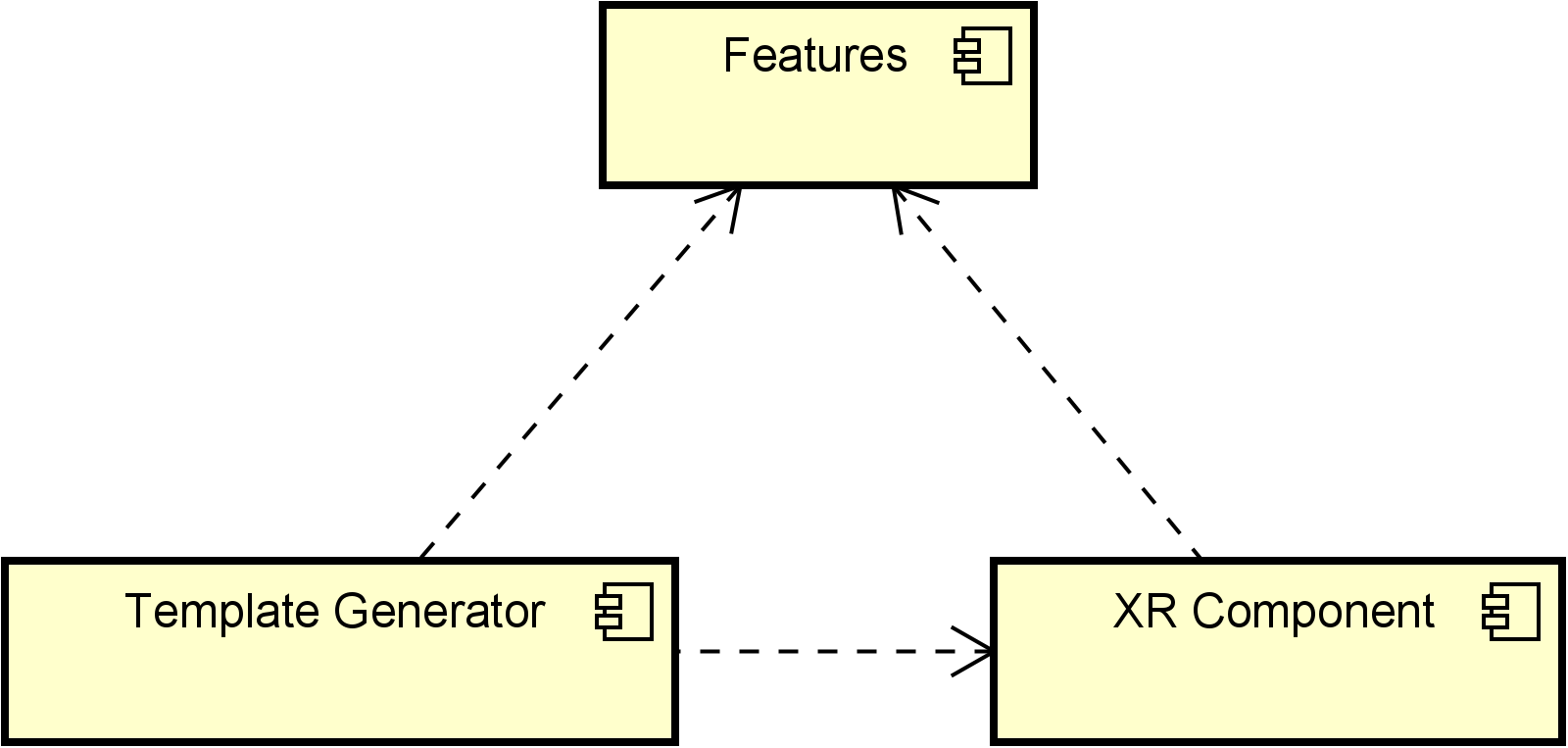}
    \caption{MetaSEE Features Model Editor components}
    \label{fig:components}
\end{figure}

Fig. \ref{fig:components} shows the dependency relationship between the components. The generated code skeleton must conform to the features as well as the A-Frame components.

\begin{figure*}[ht]
    \centering
    \includegraphics[scale=1.2]{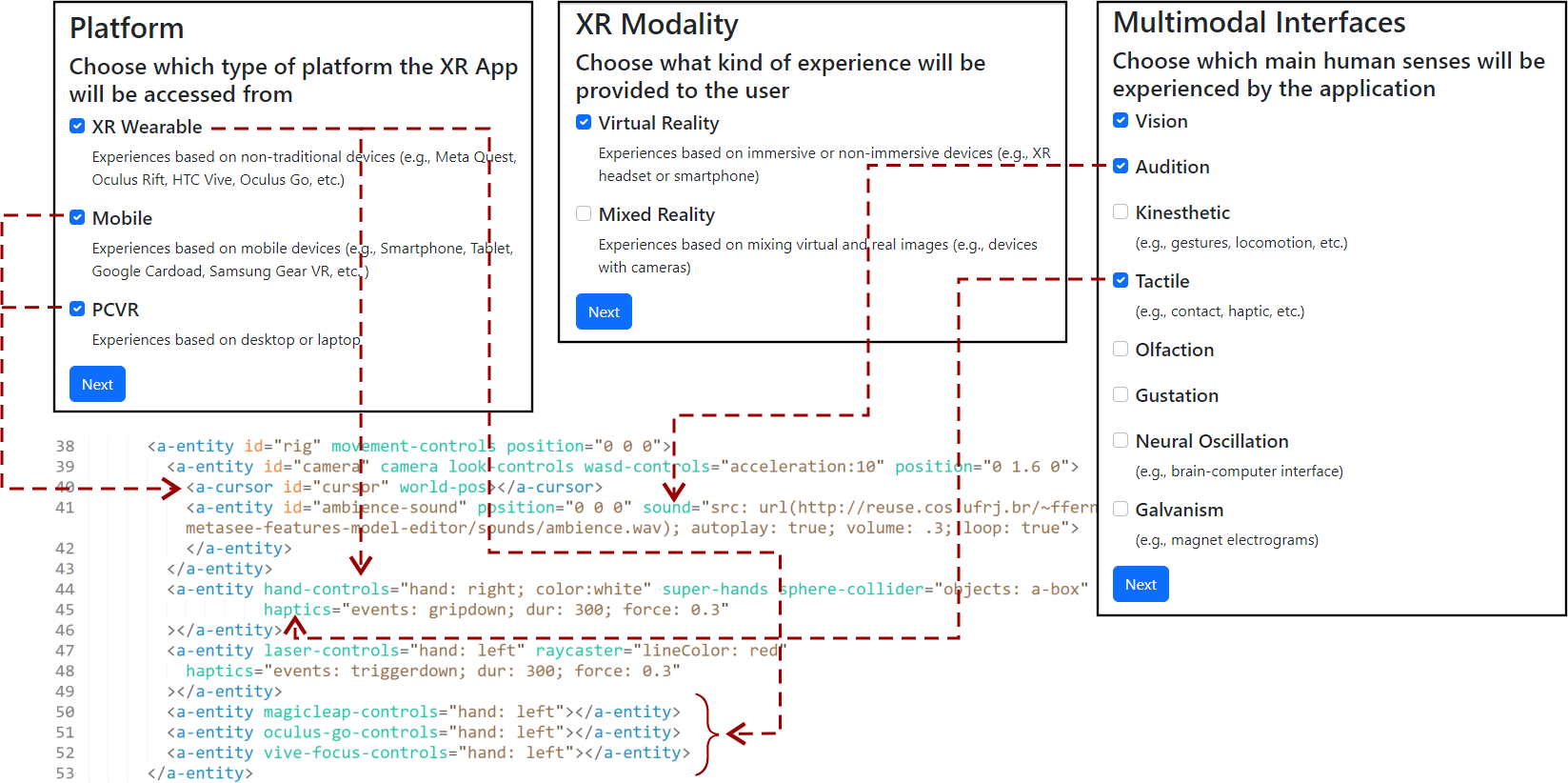}
    \caption{Relationship between features and generated source code}
    \label{fig:screens}
\end{figure*}

Fig. \ref{fig:screens} shows the relationship between the features and the generated source code. The editor's web interface presents the features that must be configured. The user on each screen must select which features will compose his/her application. In this example, the Web XR app must work on wearable, mobile, and PCVR platforms; the virtual world will consist only of virtual objects (VR); and the vision, audition, and tactile senses will be used during the immersive experience.

The piece of code is a cut of the source code skeleton generated. Considering that the three platforms were selected, the application must ensure the different forms of interaction. For example, if the application is accessed by a mobile device or desktop/laptop, the \textit{a-cursor} entity is implemented. It allows basic interactivity with a scene. The standard appearance is a ring geometry. If accessed by an XR headset (Wearable), the tracked-controls component are implemented. According to the piece of code, line 44 implements the right hand and lines 50, 51 and 52 implement the left hand.

\begin{figure}[ht]
    \centering
    \includegraphics[scale=.35]{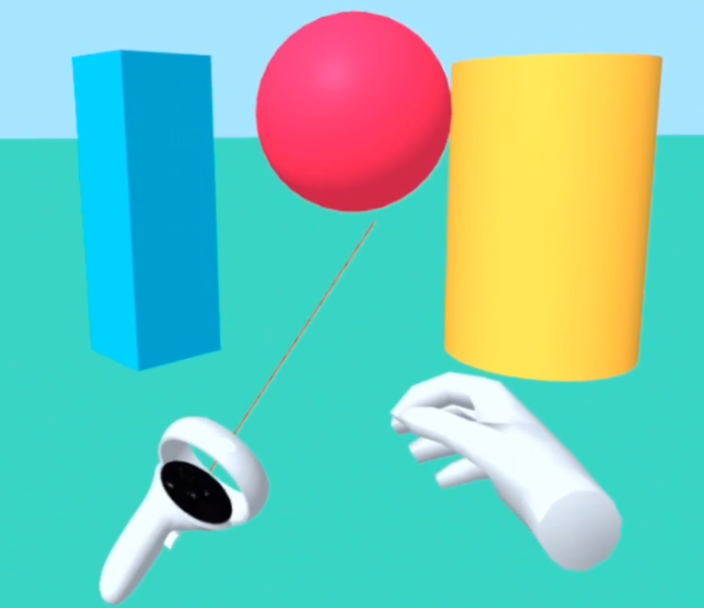}
    \caption{Application generated through the MetaSEE Features Model Editor}
    \label{fig:app-screen}
\end{figure}

According to Fig. \ref{fig:app-screen}, the left hand is represented in the virtual world by a control and the right hand by a virtual hand. The difference between the two implementations is that \textit{hand-controls} is device independent, and for each type of XR headset a component is implemented (\textit{magicleap-controls}, \textit{oculus-go-controls}, \textit{live-focus-controls}).

The virtual representation of each hand was based on the selection of the \textit{tactile} feature. By default, the code implements the \textit{haptics} component with the \textit{gripdown} event. A-Frame documentation \cite{a-frame_-frame_2022} lists other events that can be used. \textit{Gripdown} is the hand is closed into a fist without thumb raised. The \textit{dur} property defines the control vibration time in milliseconds, and \textit{force} is the vibration intensity. Therefore, the virtual hand induces the user ``to grab" the virtual objects, and the left hand (virtual control) with the actions of walking and clicking. Finally, the component \textit {sound} implements the hearing in the application.

%% file: Sections/PreliminaryEvaluation.tex
\section{Preliminary Evaluation} \label{sec:evaluation}
A preliminary evaluation of the MetaSEE Features Model Editor prototype was performed. The purpose of the evaluation is to validate the features model as well as to verify the usability of the interface.

The evaluation was carried out asynchronously during the month of September 2022. The form link was made available on several communication channels, such as Discord servers, LinkedIn, and email list. In total, 5 people evaluated the prototype. From a profile point of view, the average contact of participants with XR apps is 3 years. Most of them (3) declared as XR developers, and 2 as XR creator/designer.

Regarding the prototype interface, participants answered 5 points of Likert scale. From the point of view if the interface is friendly, 2 participants agree, 2 are neutral and 1 disagree. As for ease of navigation, 2 participants agree strongly, 2 agree, and 1 is neutral. Regarding developmental support potential, 3 participants agree and 2 disagree strongly. It was also asked if the chosen features were implemented. A participant disagrees strongly and 4 strongly agree.

Finally, an open question allowed the participants to inform what are the opportunities for improvement of the prototype. In general, the selection of more items such as screen elements and forms of interaction were suggested. In addition, it was also suggested as an improvement that the prototype has a VR interface, as well as object inventory management.

Although only 5 participants evaluated the prototype, we believe that the results are promising. We consider that the mapped feature model (see Fig. \ref{fig:features-model}) has been validated, as no participant reported any conceptual nonconformity during the selection of features through the screens (see Fig. \ref{fig:screens} ).

In addition, we believe that access to the skeleton of the generated application code could positively influence the understanding of the prototype potential to support the development of Web XR apps. Additionally, obtaining participants' information regarding the domain of programming languages and XR frameworks, especially A-Frame, would help us better understand the results. If the participant does not know how to implement A-Frame applications, he/she is unaware of the complexity of ensuring that the application works and ensures the interaction on three device platforms, for example.

%% file: Sections/Conclusion.tex
\section{Conclusion and Future Directions} \label{sec:conclusion}
This work aimed to present a SPL-based approach to support the instantiating of a Web XR app family. The approach consists of three phases: (i) domain analysis defines a model features that provides for similarities and variabilities, especially from the multimodal interactions of the applications; (ii) domain project defines a set of generic and reusable software artifacts; and (iii) domain implementation defines a MetaSEE Features Model Editor interface to generate the code skeleton of a Web XR app family from the artifacts built in the previous phases.

A preliminary evaluation with 5 participants was conducted in order to validate the features model. Despite the small number of participants, we could see that the features model can represent the complexity of the multimodal interaction of Web XR apps.

As future work, we intend to increase the features model with other elements, such as User Interface (UI) for XR and multiple avatars. Consequently, the generic and reusable artifacts of the domain project as well as the MetaSEE Features Model Editor interface will be iterated and incremented. After these improvements, an evaluation with a significant number of participants will be conducted in order to validate the approach as a whole.

%% file: Paper.bbl
\begin{thebibliography}{10}
\providecommand{\url}[1]{#1}
\csname url@samestyle\endcsname
\providecommand{\newblock}{\relax}
\providecommand{\bibinfo}[2]{#2}
\providecommand{\BIBentrySTDinterwordspacing}{\spaceskip=0pt\relax}
\providecommand{\BIBentryALTinterwordstretchfactor}{4}
\providecommand{\BIBentryALTinterwordspacing}{\spaceskip=\fontdimen2\font plus
\BIBentryALTinterwordstretchfactor\fontdimen3\font minus
  \fontdimen4\font\relax}
\providecommand{\BIBforeignlanguage}[2]{{%
\expandafter\ifx\csname l@#1\endcsname\relax
\typeout{** WARNING: IEEEtran.bst: No hyphenation pattern has been}%
\typeout{** loaded for the language `#1'. Using the pattern for}%
\typeout{** the default language instead.}%
\else
\language=\csname l@#1\endcsname
\fi
#2}}
\providecommand{\BIBdecl}{\relax}
\BIBdecl

\bibitem{wang_survey_2022}
\BIBentryALTinterwordspacing
Y.~Wang, Z.~Su, N.~Zhang, R.~Xing, D.~Liu, T.~H. Luan, and X.~Shen, ``A
  {Survey} on {Metaverse}: {Fundamentals}, {Security}, and {Privacy},''
  \emph{IEEE Communications Surveys \& Tutorials}, pp. 1--1, 2022,
  arXiv:2203.02662 [cs]. [Online]. Available:
  \url{http://arxiv.org/abs/2203.02662}
\BIBentrySTDinterwordspacing

\bibitem{ball_metaverse_2022}
M.~Ball, \emph{\BIBforeignlanguage{Inglês}{The {Metaverse}: {And} {How} {It}
  {Will} {Revolutionize} {Everything}}}.\hskip 1em plus 0.5em minus 0.4em\relax
  New York, NY: Liveright Publishing Corporation, Jul. 2022.

\bibitem{fernandes_towards_2022}
\BIBentryALTinterwordspacing
F.~Fernandes, ``Towards {Metaverse} {Engineering},'' \emph{Preprint}, Aug.
  2022. [Online]. Available:
  \url{http://dx.doi.org/10.13140/RG.2.2.29752.21765}
\BIBentrySTDinterwordspacing

\bibitem{figueroa_intml_2009}
\BIBentryALTinterwordspacing
P.~Figueroa, ``{InTml}: a case study on virtual reality development,'' in
  \emph{Proceedings of the 24th {ACM} {SIGPLAN} conference companion on
  {Object} oriented programming systems languages and applications}, ser.
  {OOPSLA} '09.\hskip 1em plus 0.5em minus 0.4em\relax New York, NY, USA:
  Association for Computing Machinery, Oct. 2009, pp. 745--746. [Online].
  Available: \url{http://doi.org/10.1145/1639950.1639994}
\BIBentrySTDinterwordspacing

\bibitem{gill_aframe_2017}
\BIBentryALTinterwordspacing
A.~Gill, ``{AFrame}: {A} {Domain} {Specific} {Language} for {Virtual}
  {Reality}: {Extended} {Abstract},'' in \emph{Proceedings of the 2nd
  {International} {Workshop} on {Real} {World} {Domain} {Specific}
  {Languages}}, ser. {RWDSL17}.\hskip 1em plus 0.5em minus 0.4em\relax New
  York, NY, USA: Association for Computing Machinery, Feb. 2017, p.~1.
  [Online]. Available: \url{http://doi.org/10.1145/3039895.3039899}
\BIBentrySTDinterwordspacing

\bibitem{fernandes_systematic_2022}
\BIBentryALTinterwordspacing
F.~Fernandes and C.~Werner, ``A {Systematic} {Literature} {Review} of the
  {Metaverse} for {Software} {Engineering} {Education}: {Overview},
  {Challenges} and {Opportunities},'' \emph{Preprint}, Sep. 2022. [Online].
  Available: \url{10.13140/RG.2.2.25341.64489}
\BIBentrySTDinterwordspacing

\bibitem{northrop_seis_2002}
L.~Northrop, ``{SEI}'s software product line tenets,'' \emph{IEEE Software},
  vol.~19, no.~4, pp. 32--40, Jul. 2002, conference Name: IEEE Software.

\bibitem{clements_software_2002}
P.~Clements and L.~Northrop, \emph{Software product lines}.\hskip 1em plus
  0.5em minus 0.4em\relax Addison-Wesley Boston, 2002.

\bibitem{fernandes_feature_2008}
P.~Fernandes, C.~Werner, and L.~Murta, ``Feature {Modeling} for
  {Context}-{Aware} {Software} {Product} {Lines},'' in \emph{Proceedings of the
  20th {International} {Conference} on {Software} {Engineering} \& {Knowledge}
  {Engineering} ({SEKE})}, San Francisco, CA, USA, 2008, pp. 758--763.

\bibitem{zaina_uma_2015}
L.~Zaina, A.~Leles, A.~Duarte, G.~Góis, and E.~F. Welter, ``Uma linha de
  produto de software para construçao de museus virtuais para aprendizagem,''
  in \emph{Brazilian {Symposium} on {Computers} in {Education} ({Simpósio}
  {Brasileiro} de {Informática} na {Educação}-{SBIE})}, vol.~26, 2015,
  p.~51, issue: 1.

\bibitem{gottschalk_proconar_2020}
S.~Gottschalk, E.~Yigitbas, E.~Schmidt, and G.~Engels,
  ``\BIBforeignlanguage{en}{{ProConAR}: {A} {Tool} {Support} for
  {Model}-{Based} {AR} {Product} {Configuration}},'' in
  \emph{\BIBforeignlanguage{en}{Human-{Centered} {Software} {Engineering}}},
  ser. Lecture {Notes} in {Computer} {Science}, R.~Bernhaupt, C.~Ardito, and
  S.~Sauer, Eds.\hskip 1em plus 0.5em minus 0.4em\relax Cham: Springer
  International Publishing, 2020, pp. 207--215.

\bibitem{lebiedz_cave_2021}
\BIBentryALTinterwordspacing
J.~Lebiedz and B.~Wiszniewski, ``{CAVE} applications: from craft manufacturing
  to product line engineering,'' in \emph{Proceedings of the 27th {ACM}
  {Symposium} on {Virtual} {Reality} {Software} and {Technology}}, ser. {VRST}
  '21.\hskip 1em plus 0.5em minus 0.4em\relax New York, NY, USA: Association
  for Computing Machinery, Dec. 2021, pp. 1--2. [Online]. Available:
  \url{http://doi.org/10.1145/3489849.3489948}
\BIBentrySTDinterwordspacing

\bibitem{figueroa_intml_2008}
P.~Figueroa, W.~F. Bischof, P.~Boulanger, H.~J. Hoover, and R.~Taylor,
  ``{InTml}: {A} {Dataflow} {Oriented} {Development} {System} for {Virtual}
  {Reality} {Applications},'' \emph{Presence}, vol.~17, no.~5, pp. 492--511,
  Oct. 2008, conference Name: Presence.

\bibitem{dionisio_3d_2013}
\BIBentryALTinterwordspacing
J.~D.~N. Dionisio, W.~G.~B. III, and R.~Gilbert, ``{3D} {Virtual} {Worlds} and
  the {Metaverse}: {Current} {Status} and {Future} {Possibilities},'' \emph{ACM
  Comput. Surv.}, vol.~45, no.~3, Jul. 2013, place: New York, NY, USA
  Publisher: Association for Computing Machinery. [Online]. Available:
  \url{https://doi-org.ez29.periodicos.capes.gov.br/10.1145/2480741.2480751}
\BIBentrySTDinterwordspacing

\bibitem{blois_variability_2006}
A.~P. T.~B. Blois, R.~F. de~Oliveira, N.~Maia, C.~Werner, and K.~Becker,
  ``\BIBforeignlanguage{en}{Variability {Modeling} in a {Component}-{Based}
  {Domain} {Engineering} {Process}},'' in \emph{\BIBforeignlanguage{en}{Reuse
  of {Off}-the-{Shelf} {Components}}}, ser. Lecture {Notes} in {Computer}
  {Science}, M.~Morisio, Ed.\hskip 1em plus 0.5em minus 0.4em\relax Berlin,
  Heidelberg: Springer, 2006, pp. 395--398.

\bibitem{braga_odyssey_1999}
R.~Braga, C.~Werner, and M.~Mattoso, ``Odyssey: a reuse environment based on
  domain models,'' in \emph{Proceedings 1999 {IEEE} {Symposium} on
  {Application}-{Specific} {Systems} and {Software} {Engineering} and
  {Technology}. {ASSET}'99 ({Cat}. {No}.{PR00122})}, Mar. 1999, pp. 50--57.

\bibitem{kang_feature-oriented_2002}
K.~Kang, J.~Lee, and P.~Donohoe, ``Feature-oriented product line engineering,''
  \emph{IEEE Software}, vol.~19, no.~4, pp. 58--65, Jul. 2002, conference Name:
  IEEE Software.

\bibitem{augstein_human-centered_2019}
\BIBentryALTinterwordspacing
M.~Augstein and T.~Neumayr, ``A {Human}-{Centered} {Taxonomy} of {Interaction}
  {Modalities} and {Devices},'' \emph{Interacting with Computers}, vol.~31,
  no.~1, pp. 27--58, Jan. 2019. [Online]. Available:
  \url{https://doi.org/10.1093/iwc/iwz003}
\BIBentrySTDinterwordspacing

\bibitem{milgram_augmented_1995}
\BIBentryALTinterwordspacing
P.~Milgram, H.~Takemura, A.~Utsumi, and F.~Kishino, ``Augmented reality: a
  class of displays on the reality-virtuality continuum,'' in
  \emph{Telemanipulator and {Telepresence} {Technologies}}, vol. 2351.\hskip
  1em plus 0.5em minus 0.4em\relax SPIE, Dec. 1995, pp. 282--292. [Online].
  Available:
  \url{https://www-spiedigitallibrary.ez29.periodicos.capes.gov.br/conference-proceedings-of-spie/2351/0000/Augmented-reality--a-class-of-displays-on-the-reality/10.1117/12.197321.full}
\BIBentrySTDinterwordspacing

\bibitem{a-frame_-frame_2022}
\BIBentryALTinterwordspacing
{A-Frame}, ``\BIBforeignlanguage{en}{A-{Frame}},'' Sep. 2022. [Online].
  Available: \url{https://aframe.io}
\BIBentrySTDinterwordspacing

\bibitem{gamma_design_1995}
E.~Gamma, R.~Helm, R.~Johnson, R.~E. Johnson, and J.~Vlissides,
  \emph{\BIBforeignlanguage{de}{Design {Patterns}: {Elements} of {Reusable}
  {Object}-{Oriented} {Software}}}.\hskip 1em plus 0.5em minus 0.4em\relax
  Pearson Deutschland GmbH, 1995, google-Books-ID: tmNNfSkfTlcC.

\end{thebibliography}
